\documentclass[5p]{elsarticle}
\usepackage{amsmath}
\usepackage{amsfonts}
\usepackage{graphicx}
\usepackage{subcaption}
\usepackage{tkz-euclide}
\usetkzobj{all}
\usepackage{graphicx} 
\usepackage[colorlinks=true,hyperfootnotes=true,breaklinks=true]{hyperref}

\DeclareMathOperator{\sign}{\mathrm{sign}}

\begin{document}

\begin{frontmatter}
\title{{Towards the Heider balance with a cellular automaton}}

\author{Krzysztof Malarz\corref{km}\fnref{orcidkm}}  
\ead{malarz@agh.edu.pl}  
\author{Maciej Wo{\l}oszyn\fnref{orcidmw}}
\ead{woloszyn@agh.edu.pl}
\author{Krzysztof Ku{\l}akowski\fnref{orcidkk}}
\ead{kulakowski@fis.agh.edu.pl}

\address{AGH University of Science and Technology,
Faculty of Physics and Applied Computer Science,\\
al. Mickiewicza 30, 30-059 Kraków, Poland.}

\cortext[km]{Corresponding author}
\fntext[orcidkm]{\includegraphics[scale=0.06]{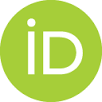} \href{http://orcid.org/0000-0001-9980-0363}{0000-0001-9980-0363}}
\fntext[orcidmw]{\includegraphics[scale=0.06]{ORCID} \href{http://orcid.org/0000-0001-9896-1018}{0000-0001-9896-1018}}
\fntext[orcidkk]{\includegraphics[scale=0.06]{ORCID} \href{http://orcid.org/0000-0003-1168-7883}{0000-0003-1168-7883}}

\begin{abstract}
The state of structural balance (termed also `Heider balance') of a social network is often discussed in social psychology and sociophysics. In this state, actors at network nodes classify other individuals as enemies or friends. Hence, the network contains two kinds of links: positive and negative. Here a new cellular automaton is designed and investigated, which mimics the time evolution towards the structural balance. The automaton is deterministic and synchronous. The medium is the triangular lattice with some fraction $f$ of links removed. We analyse the number of unbalanced triads (parameterized as `energy'), the frequencies of balanced triads and correlations between them. The time evolution enhances the local correlations of balanced triads. Local configurations of unbalanced triads are found which are blinking with period of two time steps.  
\end{abstract}

\end{frontmatter}

\section{Introduction}

\begin{figure}[htbp]
\begin{subfigure}{.240\columnwidth}
        \caption{\label{1a}}
        \centering
\begin{tikzpicture}[scale=1.2]
\draw[blue,very thick] (0,0) -- (1,0);
\node[below] at (0.5, 0) {$+$};
\draw[blue,very thick] (0,0) -- (0.5,0.866);
\node[left] at (0.25, 0.433) {$+$};
\draw[blue,very thick] (1,0) -- (0.5,0.866);
\node[right] at (0.75, 0.433) {$+$};
\end{tikzpicture}
        \end{subfigure}
        \begin{subfigure}{.240\columnwidth}
        \caption{\label{1b}}
        \centering
\begin{tikzpicture}[scale=1.0]
\draw[blue,very thick] (2,0) -- (3,0);
\node[below] at (2.5, 0) {$+$};
\draw[blue,very thick] (2,0) -- (2.5,0.866);
\node[left] at (2.25, 0.433) {$+$};
\draw[red,very thick, dashed]  (3,0) -- (2.5,0.866);
\node[right] at (2.75, 0.433) {$-$};
\end{tikzpicture}
        \end{subfigure}
        \begin{subfigure}{.240\columnwidth}
        \caption{\label{1c}}
        \centering
\begin{tikzpicture}[scale=1.0]
\draw[blue,very thick] (4,0) -- (5,0);
\node[below] at (4.5, 0) {$+$};
\draw[red,very thick, dashed]  (4,0) -- (4.5,0.866);
\node[left] at (4.25, 0.433) {$-$};
\draw[red,very thick, dashed]  (5,0) -- (4.5,0.866);
\node[right] at (4.75, 0.433) {$-$};
\end{tikzpicture}
        \end{subfigure}
        \begin{subfigure}{.240\columnwidth}
        \caption{\label{1d}}
        \centering
\begin{tikzpicture}[scale=1.2]
\draw[red,very thick, dashed]  (6,0) -- (7,0);
\node[below] at (6.5, 0) {$-$};
\draw[red,very thick, dashed]  (6,0) -- (6.5,0.866);
\node[left] at (6.25, 0.433) {$-$};
\draw[red,very thick, dashed]  (7,0) -- (6.5,0.866);
\node[right] at (6.75, 0.433) {$-$};
\end{tikzpicture}
        \end{subfigure}
\caption{\label{Heidertriads} Heider's triads corresponding to balanced (the first and the third from the left) and imbalanced (the second and the fourth) states. Continuous blue lines and dashed red lines represent friendly and hostile relations, respectively.}
\end{figure}
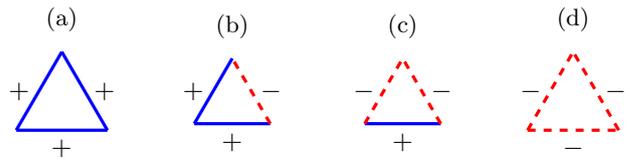

The problem of structural balance \cite{Heider,Cartwright,Harary} is one of key mathematical contributions to sociology \cite{Bonacich}. 
In a balanced state of a complete signed network, where links are either positive or negative, the product of links of each triad is 
positive.
Consequently, the network is divided in two groups, with positive links within each part and negative links between nodes in different parts.
The concept of structural or Heider balance has been often discussed in terms of removal of cognitive dissonance \cite{Festinger,Brashears}.
{Consider a triad of actors, A, B and C, placed at the nodes of a triangle, with friendly or hostile relations 
between them. These kinds of relations are encoded as signs of the links, positive or negative ones. Suppose that the links
AB and BC are positive, and AC is negative. Then A suffers a dissonance: why his friend B is a friend of C, who is enemy of A?
The solution is either to make a friendly relation with C, or to quarrel with B. Then, the dissonance is removed; the balance is restored.
Balanced states (presented in Figs.}~\ref{1a} {and} \ref{1c}{) are usually characterized with the following `algebra'} \cite{Aronson}:
\begin{enumerate}
\item {friend of my friend is my friend,}
\item {friend of my enemy is my enemy,} 
\item {enemy of my friend is my enemy,} 
\item {enemy of my enemy is my friend.}
\end{enumerate}

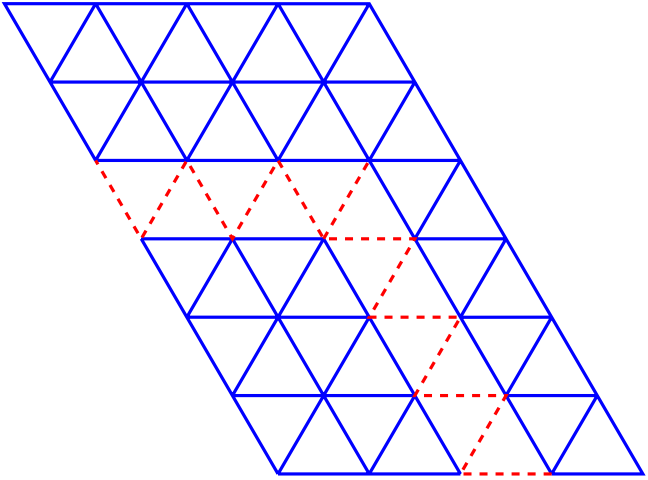
\begin{figure}
\begin{tikzpicture}[scale=1.2]
\draw[blue,very thick]
( 0.0,0.000) -- ( 2.0,0.000)
( 3.0,0.000) -- ( 4.0,0.000) -- (1.0,5.195) -- (-3.0,5.196) -- (-2.0,3.464)
(-1.5,2.598) -- ( 0.0,0.000)
( 1.0,0.000) -- (-0.5,2.598) (-1.0,3.464) -- (-2.0,5.196)
( 2.0,0.000) -- ( 0.5,2.598) ( 0.0,3.464) -- (-1.0,5.196)
(-1.5,2.598) -- ( 0.5,2.598) ( 1.5,2.598) -- ( 2.5,2.598)
(-2.0,3.464) -- ( 2.0,3.464)
(-2.5,4.330) -- ( 1.5,4.330)
(-0.5,0.866) -- ( 1.5,0.866) ( 2.5,0.866) -- ( 3.5,0.866)
(-1.0,1.732) -- ( 1.0,1.731) ( 2.0,1.731) -- ( 3.0,1.732)

(-2.5,4.330) -- (-2.0,5.196)
(-2.0,3.464) -- (-1.0,5.196)
(-1.0,3.464) -- ( 0.0,5.196)
(-1.0,1.732) -- (-0.5,2.598) ( 0.0,3.464) -- ( 1.0,5.196)
(-0.5,0.866) -- ( 0.5,2.598) ( 1.0,3.464) -- ( 1.5,4.330)
( 0.0,0.000) -- ( 1.0,1.732) ( 1.5,2.598) -- ( 2.0,3.464)
( 1.0,0.000) -- ( 1.5,0.866) ( 2.0,1.732) -- ( 2.5,2.598)
( 2.5,0.866) -- ( 3.0,1.732)
( 3.0,0.000) -- ( 3.5,0.866)
( 3.0,0.000) -- ( 0.0,5.196);

\draw[red,very thick, dashed] (3,0) -- (2,0) -- (2.5,0.866) -- (1.5,0.866) -- (2,1.732) -- (1,1.732) -- (1.5,2.598) -- (0.5,2.598) -- (0,3.464) -- (-0.5,2.598) -- (-1,3.464) -- (-1.5,2.598) -- (-2,3.464);
\draw[red,very thick, dashed] (0.5,2.598) -- (1,3.464);

\end{tikzpicture}
\caption{\label{lattice}{A boundary between two internally friendly groups of nodes, with hostile intergroup relations. All triads are balanced.}}
\end{figure}

{As follows from these rules, the state with all links positive is balanced (the so-called `paradise'), while the state with all links 
negative is not balanced and therefore unstable under the considered dynamics. Also, positive relations are transitive (friend of my 
friend is my friend) while negative relations are not (enemy of my enemy is my friend).

It was proved by Cartwright and Harary} \cite{Cartwright}{, that a complete network with all triads balanced is equivalent to a division 
of the system into two internally friendly, but mutually hostile groups. Here we discuss the case of triangular lattice, where 
links exist only between nearest neighbors. However, it is easy to construct a similar division of the lattice in the balanced
state. An example of such a state is shown in Fig.}~\ref{lattice}{. There is a clear boundary between two groups, with all triads balanced.
The shape of the boundary can be modified; this means that the balanced state is not unique. However, as will be shown below, 
the balance is not generic in the sense that it does not appear as an outcome of the evolution from a random initial state. 
Rather, the system is stuck in one of numerous metastable states, where energy is not minimal.

It is straightforward to notice that balanced states exist in a network with any topology. One has to take a balanced state in a complete graph
and to remove links as to transform the graph to the desired network. The balance cannot be destroyed by removal of links. Also, the 
initial division cannot be changed, if only the network remains connected.
}

Several algorithms have been formulated to reproduce the process of gaining the balance \cite{Antal,Kulakowski2005,Antal-2006130,Marvel,Gorski,Krawczyk,Kulakowski-2019,1911.13048}. 
In all these cases, interaction between a pair of nodes does depend on the states of links from these nodes to all other nodes.
In this sense, the interaction is non-local. The drawback of this approach is that the considerations are limited to complete graphs.
Our aim here is to explore an evolution towards balance, determined by interaction with a few nearest neighbours. The time evolution 
of the system is given by a specially designed cellular automaton, which is local by definition. The automaton rule mimics the differential 
equation of motion \cite{Kulakowski2005}, which has been proved to lead to the balance in the global version.

In the next section we specify the automaton rule. Numerical results are reported in Section~\ref{sec-res}. In Section~\ref{sec-mod} approximate formulas are provided on a mean state of a complete triad. Discussion and summary are given in last section.

\section{The rule}

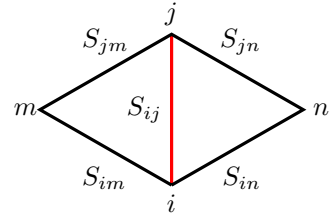
\begin{figure}[htbp]
	\centering
\begin{tikzpicture}[scale=2.0]
	\draw[red,very thick]  (0,0)--(0,1);
	\draw[very thick] (0,0)--(-.866,.5)--(0,1)--(.866,.5)--(0,0);
	\node[below] at (0,0) {$i$};
	\node[above] at (0,1) {$j$};
	\node[right] at (-1.1,.5) {$m$};
	\node[left] at (1.1,.5) {$n$};
	\node[left] at (0,.5) {$S_{ij}$};
	\node[left] at (.65,.05) {$S_{in}$};
	\node[left] at (.65,.95) {$S_{jn}$};
	\node[right] at (-.65,.05) {$S_{im}$};
	\node[right] at (-.65,.95) {$S_{jm}$};
\end{tikzpicture}
	\caption{\label{fig:rule}The configuration of signed links $S_{im}$, $S_{in}$, $S_{jm}$, $S_{jn}$ which influence the link $S_{ij}$ in the next time step: $S_{ij}(t+1)=\sign\big(S_{im}(t)S_{jm}(t)+S_{in}(t)S_{jn}(t)\big)$. If the value of the right side is zero, $S_{ij}(t+1)=S_{ij}(t)$.}
\end{figure}

As each link can be in two states $S_{ij}=\pm1$, there are four possible states of a triad $ijk$: two balanced where the product $S_{ij}S_{jk}S_{ki}=+1$ and two unbalanced where $S_{ij}S_{jk}S_{ki}=-1$. These triads are shown in Fig. \ref{Heidertriads}. The automaton is deterministic and synchronous. The rule is defined for a triangular lattice---an array of triads---which can be diluted: links are removed with probability $f$, which is a parameter. The network topology remains unchanged during the simulation; it is only the signs of links, positive or negative, which evolve. For $f$=0, for each link $S_{ij}$ between the nodes $i$ and $j$ four links $S_{im}$, $S_{in}$, $S_{jm}$ and $S_{jn}$ are checked to two nodes $m$, $n$ which are nearest neighbours of both $i$ and $j$. The rule is 
\begin{equation}
S_{ij}(t+1)=\sign\big(S_{im}(t)S_{jm}(t)+S_{in}(t)S_{jn}(t)\big).
\label{rule}
\end{equation}
If the argument of the function $\sign(\cdots)$ is zero, $S_{ij}$ remains unchanged. The same rule is applied for diluted lattice, where $f>0$. In particular if
{links are absent}
to both common neighbours of $i$ and $j$, the link $S_{ij}$ remains frozen. The rule is shown schematically in Fig.~\ref{fig:rule}.

This rule is motivated by the following argument, which can be easily verified by inspection. Consider the situation where the links $S_{im}$, $S_{in}$, $S_{jm}$ and $S_{jn}$ are not changed. Provided that both triads $ijm$ and $ijn$ are balanced, the link $S_{ij}$ remains unchanged. If both are unbalanced, the link $S_{ij}$ changes its sign in the next time step, and both triads get balanced. Finally, if one triad is balanced and another is not, the link $S_{ij}$ remains unchanged. Of course, in many cases the links $S_{im}$, $S_{in}$, $S_{jm}$ and $S_{jn}$ do change; yet we can expect some drift towards balance. We add that in Ref.~\cite{Kulakowski2005}, a differential equation of motion has been proposed for the links, driven with the same mechanism. There, generically the evolution led to one of balanced states.

\section{\label{sec-res}Results}

\begin{figure}[!htbp]
\begin{subfigure}{.99\columnwidth}
\caption{$f=0.1$}
	\centering
\includegraphics[width=.80\textwidth]{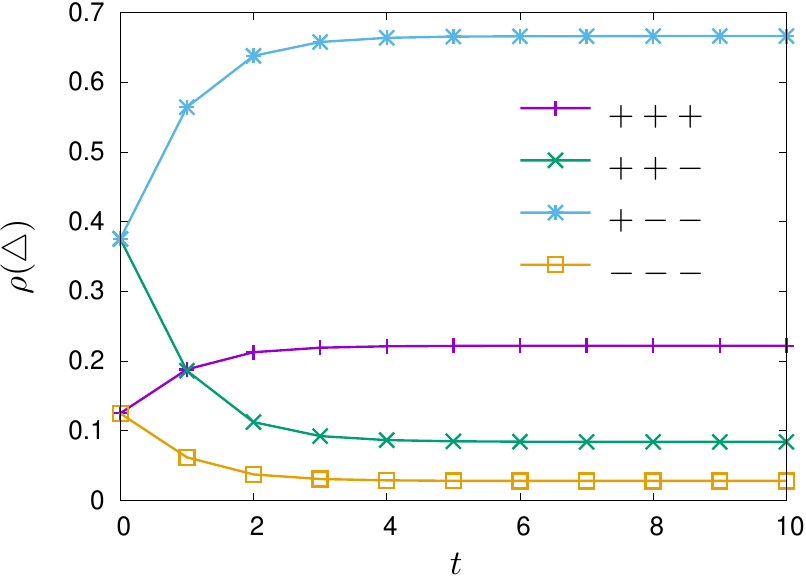}
\end{subfigure}
\begin{subfigure}{.99\columnwidth}
\caption{$f=0.5$}
	\centering
\includegraphics[width=.80\textwidth]{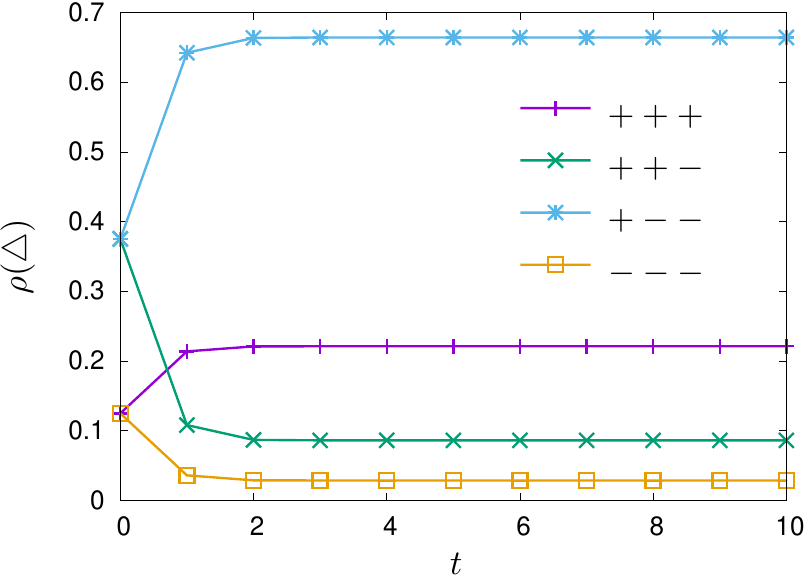}
\end{subfigure}
\begin{subfigure}{.99\columnwidth}
\caption{$f=0.9$}
	\centering
\includegraphics[width=.80\textwidth]{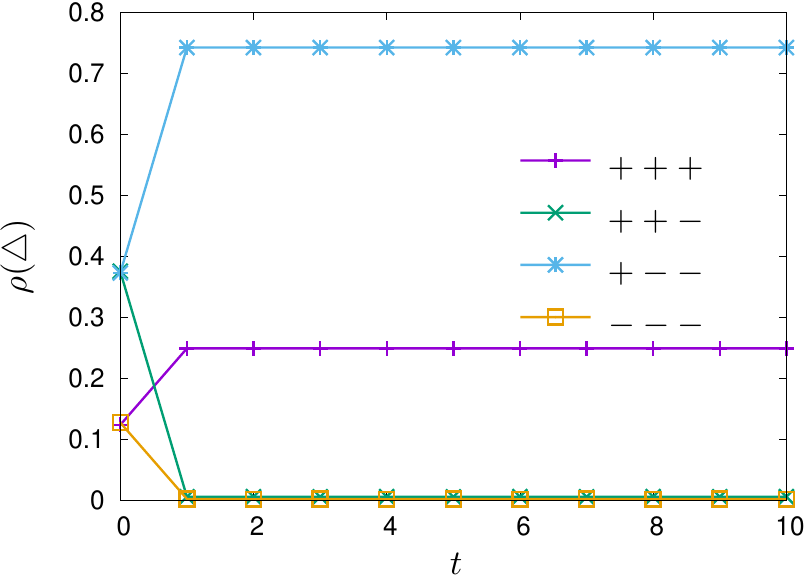}
\end{subfigure}
\caption{\label{tri}Time dependence o fractions of triads of four kinds for $f=0.1$, 0.5 and 0.9 (from top to bottom).}
\end{figure}

The calculations are performed on a triangular lattice of $N=19798$ 
 triads
 with 
 helical boundary conditions. If not stated otherwise, the results reported below are obtained with initial fraction of positive links $p$ equal to 0.5.
{The results are averaged over one hundred (for Figs.}~\ref{roro}--\ref{co}{) or one thousand (for Figs.}~\ref{tri}--
\ref{ufin}, \ref{numtria}{) simulations with various initial distribution of removed links and initial distribution of positive links.}

\subsection{Towards balance}

\begin{figure}[htbp]
	\centering
\includegraphics[width=.80\columnwidth]{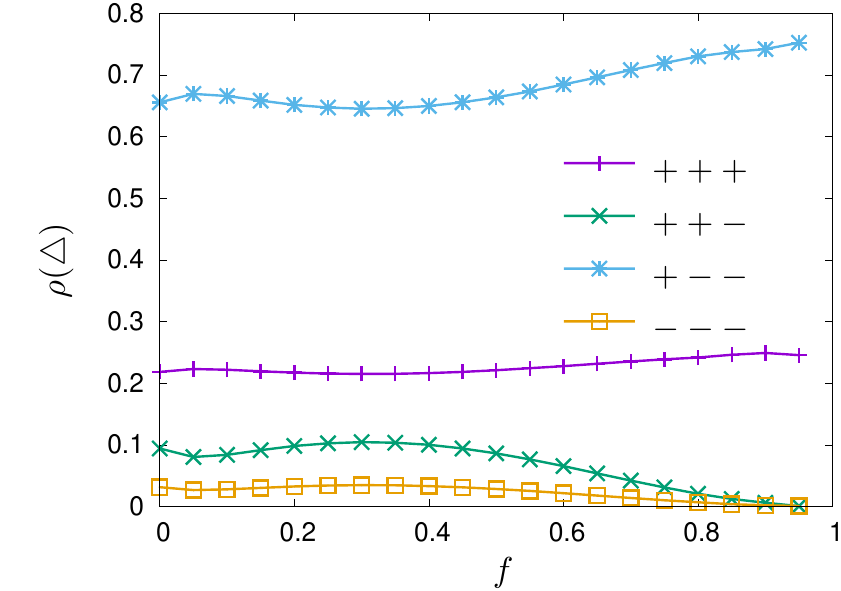}
\caption{\label{nooftra_vs_f}Fractions of triads of four kinds at final states as dependent of the dilution parameter $f$.}
\end{figure}

\begin{figure}[htbp]
\centering
\includegraphics[width=.80\columnwidth]{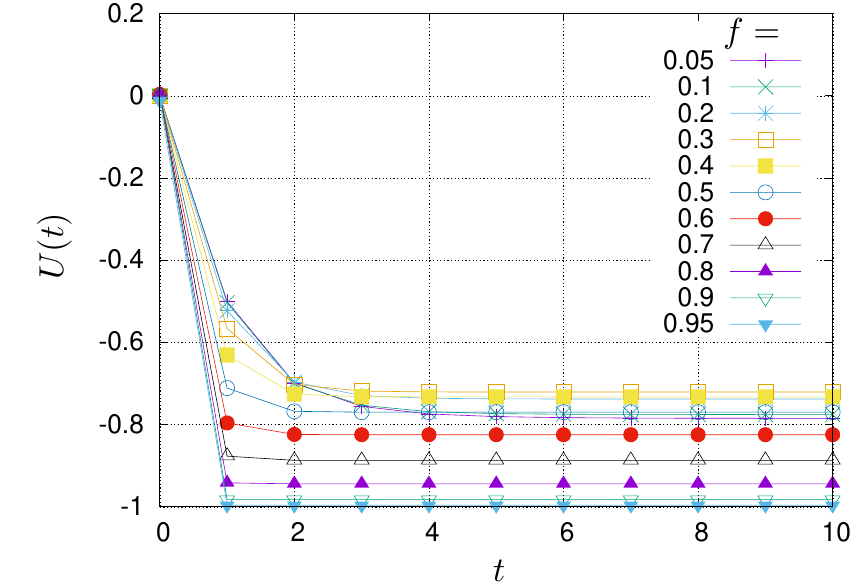}
\caption{\label{evo}Time dependence of energy $U$ for different values of the probability $f$.}
\end{figure}

Basically, the role of time evolution is to drive the system towards the balance.
In Fig.~\ref{tri}, we show that the fractions of balanced triads increase in time, and those of unbalanced---decrease.
In particular, for diluted lattice where triads are mutually isolated, unbalanced triad vanish---this effect is shown in Fig.~\ref{nooftra_vs_f}.
To measure the number of unbalanced triads in a diluted lattice, we define the sum $U$ of the products of links in a triad divided by the number of complete triads. If a link in a triad is missing, i.e. $S_{ij}=0$, this triad is not taken into account. This function $U$ is an equivalent of energy: which is to be minimised
\begin{equation}
U=-\frac{\sum_{ijk}S_{ij}S_{jk}S_{ki}}{\sum_{ijk}|S_{ij}S_{jk}S_{ki}|}.
\label{U}
\end{equation}
In a perfectly balanced state, where all triads are balanced, $U=-1$. In Fig.~\ref{evo} we show the time dependence of $U$ for different fractions $f$. In accordance with its role $U$ decreases in time and it reaches final value in less than ten time steps.

\begin{figure}[htbp]
\centering
\includegraphics[width=.80\columnwidth]{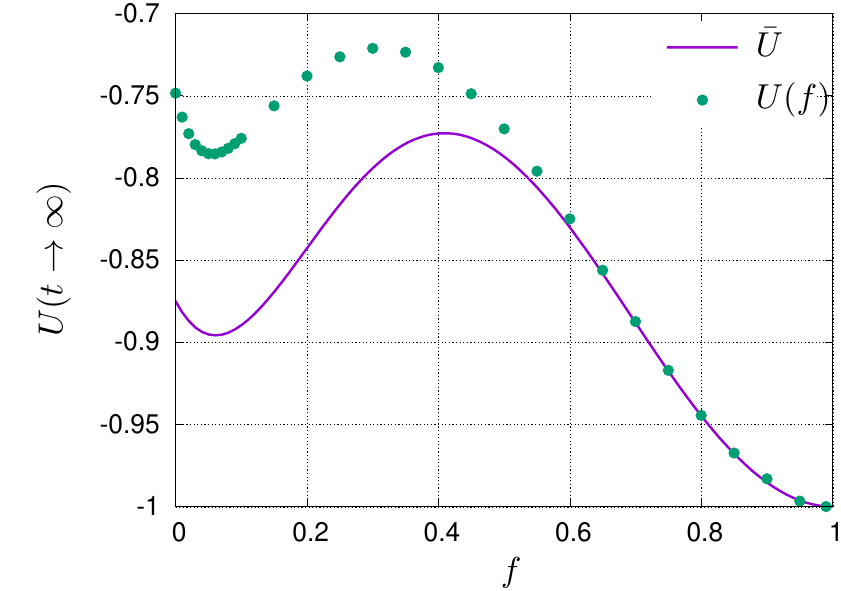}
\caption{\label{ufin}Final value of energy $U$ as dependent on the dilution parameter $f$: the results of simulations (dots) and of the model reported in Section~\ref{sec-mod} (line).}
\end{figure}

For diluted systems, this value depends on the fraction $f$ of missing links. This dependence is shown in Fig.~\ref{ufin}. For $f\approx 1$, a typical triad is isolated from other triads and therefore nothing prevents it to be balanced; hence $U=-1$. In the limit of small $f$, $U$ tends to $-3/4$, what means that in average, each eighth triad is unbalanced. For small $f$, this fraction slightly decreases with $f$, what can be interpreted as a release of some barriers, induced by interaction between triads. Such is also the explanation of a fall of energy above $f=0.4$. The maximum of $U$ near $f=0.3$
{ is reproduced within a simple approximation in Section}~\ref{sec-mod}.

\subsection{Density of negative links}

\begin{figure}[htbp]
\centering
\includegraphics[width=.80\columnwidth]{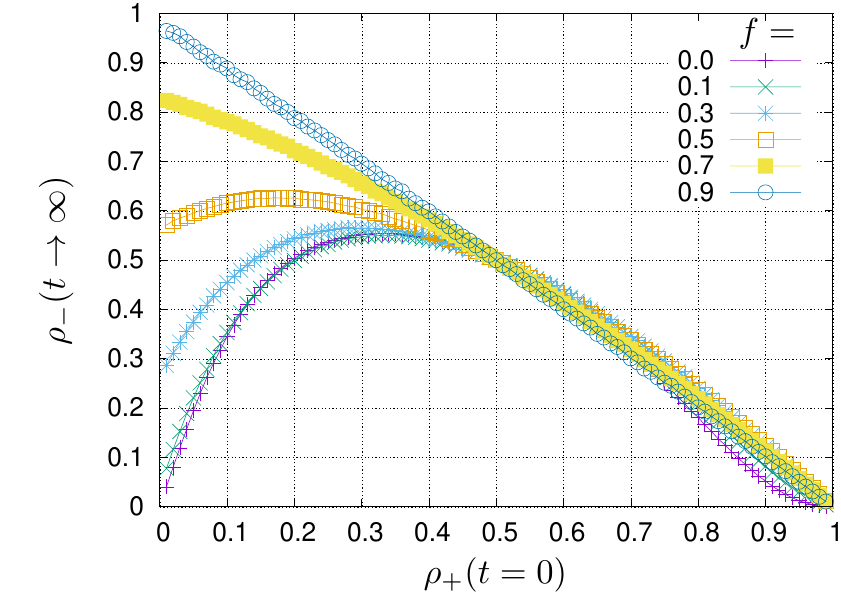}
\caption{\label{roro}Final density of negative links against initial density of positive links
for different values of the dilution parameter $f$.}
\end{figure}

When taken directly, the automaton rule does not introduce any bias towards positive or negative links. If all triads are balanced, the number of triads with all links positive is---as a rule---three times less than the other triads, what makes the numbers of positive and negative links equal. On the other hand, a phase transition has been observed in literature towards a `paradise' state, where all links are positive, if the mean value of a link exceeds some critical value \cite{Hassanibesheli}. Having this in mind, we have calculated the density 
$\rho_-(t\to\infty)$ of negative links in a final state, as dependent on the initial density $\rho_+(t=0)$ of positive links. The results are shown in Fig.~\ref{roro}. As we see, for the argument equal to 0.5 the value is also 0.5; the initial symmetry remains preserved. This remains true for all values of $f$.

\begin{figure}[htbp]
\centering
\includegraphics[width=.80\columnwidth]{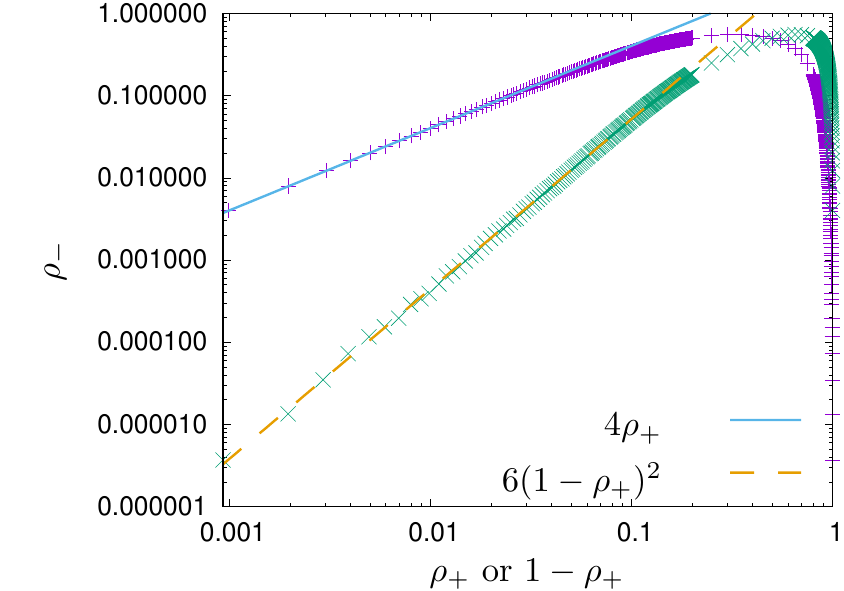}
\caption{\label{fit}Fitting of the final density of negative links for $f=0$. The original curve is given in Fig.~\ref{roro}.}
\end{figure}

What deserves a comment is a linear increase of the plot near $\rho_+(t=0)=0$ and apparently non-linear vanishing of the plot near $\rho_+(t=0)=1$. The limit values are clear: for lack of positive links at $t=0$ all links are switched to +1 at $t=1$, and this state is stable. The linearity of the plot means, that one positive link in the initial state of all other links negative produces $n$ negative links in the final state. It is easy to check by inspection that $n=4$, and such is also the slope of the blue line in Fig.~\ref{fit}. On the contrary, one negative link in the sea of positive links just vanishes. Yet, two negative links in the same triad survive in the sea of positive links; hence the parabolic dependence of $\rho_-(t\to\infty)$ on $\rho_+(t=0)$. In any case, the transition to `paradise' has not been found.

\subsection{Correlations}

\begin{figure}[htbp]
\begin{center}
\begin{subfigure}{.99\columnwidth}
\caption{$f=0.0$}
\includegraphics[width=.80\textwidth]{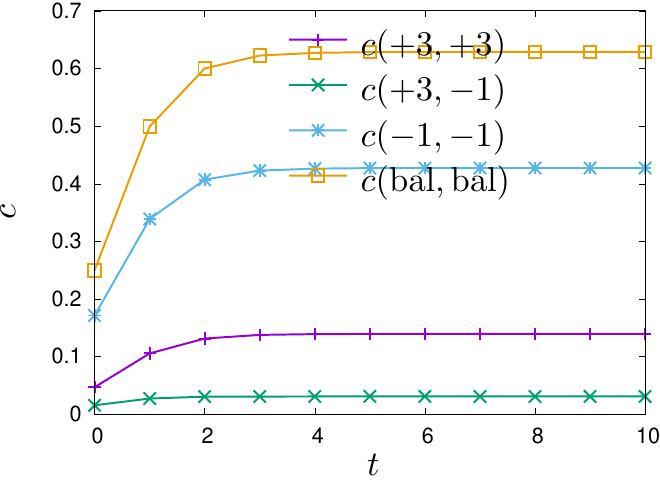}
\end{subfigure}
\begin{subfigure}{.99\columnwidth}
\caption{$f=0.2$}
\includegraphics[width=.80\textwidth]{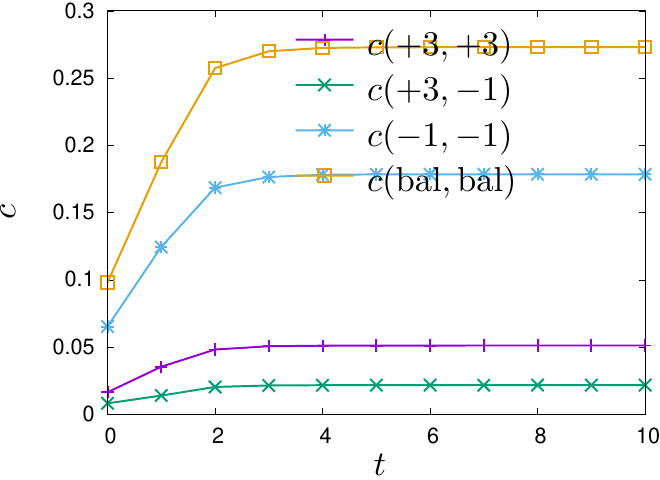}
\end{subfigure}
\begin{subfigure}{.99\columnwidth}
\caption{$f=0.5$}
\includegraphics[width=.80\textwidth]{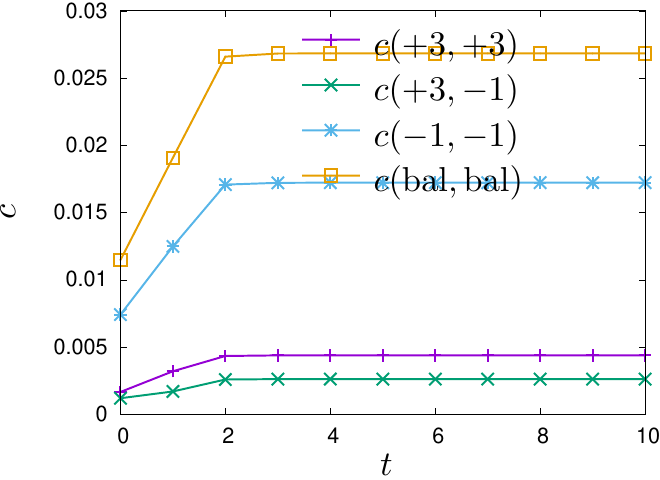}
\end{subfigure}
\begin{subfigure}{.99\columnwidth}
\caption{$f=0.8$}
\includegraphics[width=.80\textwidth]{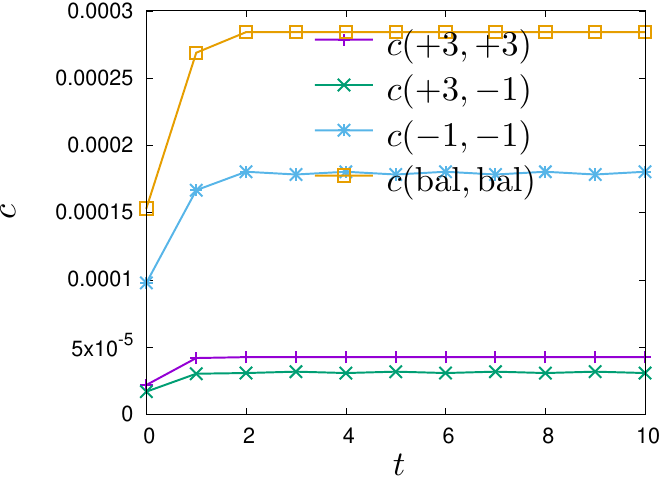}
\end{subfigure}
\end{center}
\caption{\label{co2}Correlations between different kinds of balanced triads, calculated as departures from the Bragg--Williams approximation, against time, for different values of the dilution parameter $f$.}
\end{figure}

Let us introduce a new variable $x_{ijk}=S_{ij}+S_{jk}+S_{ki}$, which allows to distinguish between different kinds of triads and is invariant under permutation of nodes $i,j,k$. We are interested in time variation of the spatial correlations of particular kinds of triads. To evaluate these correlations,
we calculate the departure of the numbers of pairs of particular neighbouring triad states from the Bragg--Williams approximation \cite[p. 513]{Huang}, \cite[p. 17]{Ziman}
\begin{subequations}
\label{cr}
\begin{equation}
	c(x,y)=\frac{N(x,y)}{3N}-\frac{N(x)N(y)}{N^2}
\end{equation}
and
\begin{equation}
	c(x,x)=\frac{2N(x,x)}{3N}-\frac{[N(x)]^2}{N^2}
\end{equation}
\end{subequations}
for $x \ne y$ and $x=y$, respectively. There, $N(x)$ is the number of triads in the state $x$, and $N(x,y)$ is the number of pairs of neighbouring triads in the states $x$ and $y$. In the triangular lattice, each triad has three neighbours.
The time dependence of these functions for $x,y$ = $+3$, $-1$ and for balanced triads is shown in Fig.~\ref{co2}.

\begin{figure}[htbp]
\centering
\includegraphics[width=.80\columnwidth]{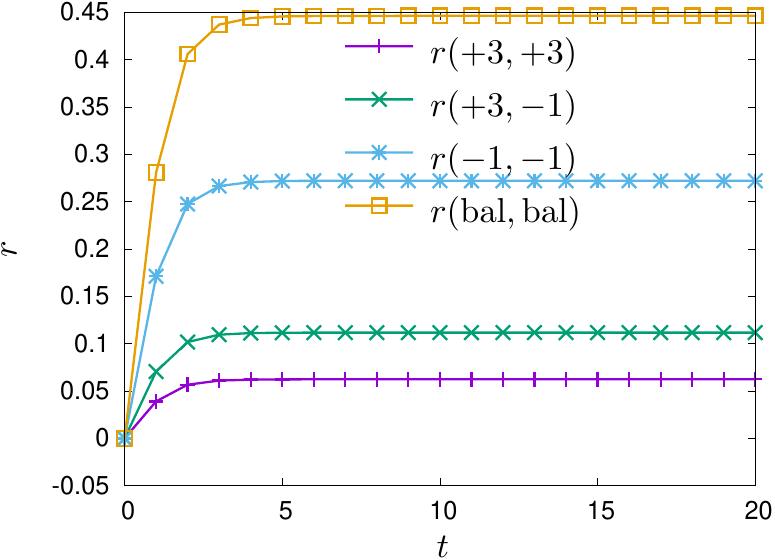}
\caption{\label{co}The time dependence of the correlations between the balanced states of triads, calculated with respect to correlations $P_0$ due to a common link of neighbouring triads. The functions $P_0$ are given in Table~\ref{corr}. The plots are obtained for $p=q=\frac{1}{2}$ and $f=0$.}
\end{figure}

In the initial state,
values of the signs of links are set randomly. Still the values of $x_{ijk}$ are correlated, because neighbouring triads share a common link. These initial correlations can be easily found by inspection. Obviously, the initial probabilities $P_0(x)$ of particular values of $x$ are: $p^3$, $3p^2(1-p)$, $3p(1-p)^2$ and $(1-p)^3$ for $x=+3$, $+1$, $-1$ and $-3$, respectively, where $p$ is the probability that a randomly selected link is positive. Further, it is easy to check by inspection that the probabilities of pairs of neighbouring balanced triads are: $P_0(+3,+3)=p^5$, $P_0(+3,-1)=2p^3q^2$ and $P_0(-1,-1)=pq^4+4p^2q^3$, where $q=1-p$. The standard formulae for the correlations are $P_0(x,y)-P_0(x)P_0(y)$ \cite{Huang}. However, here we are interested only in the contribution of the dynamics to the correlations. To evaluate this contribution, we should extract not the product $P_0(x)P_0(y)$,
but rather the correlations $P_0(x,y)$ which are related to common links of neighbouring triads. These contributions are collected in Table \ref{corr}.

Similarly, the related contribution to the correlation $r(\text{bal},\text{bal})$ of balanced triads is just the sum
\begin{equation}
	r(\text{bal},\text{bal})=r(+3,+3)+r(+3,-1)+r(-1,-1),
\label{corrgen}
\end{equation}
where $r(x,y)=P(x,y)-P_0(x,y)$, and $P(x,y)$ is the probability of a pair of neighbouring triads in any state, correlated or not. In the uncorrelated state, $P(x,y)=P_0(x,y)$. By definition, the initial value of $r(x,y)$ is zero for any value of $p$. The time dependence of the correlation functions of the balanced states is shown in Fig. \ref{co}.

Let us add that an inspection by eye of the evolving lattices shows that if two non-balanced triads are neighbours, there is some blinking there; such states are never constant. This is true for any value of the dilution coefficient $f$.

\begin{table}[htbp]
\caption{\label{corr}The probabilities $P_0(x,y)$ of $x$'s for neighbouring triads, for random signs of links; $p$ and $q=1-p$ are the probabilities of positive and negative links.}
\centering
\begin{tabular}{llc}
\hline \hline
	$x$ & $y$ & $P_0(x,y)$ \\ 
\hline \hline 
	$+3$ & $+3$	& $p^5$ \\ \hline 
	$+3$ & $+1$	& $4p^4q$ \\ \hline 
	$+3$ & $-1$	& $2p^3q^2$ \\ \hline 
	$+1$ & $+1$	& $4p^3q^2+p^4q$ \\ \hline 
	$+1$ & $-1$	& $4p^2q^2$ \\ \hline
	$+1$ & $-3$	& $2p^2q^3$ \\ \hline  
	$-1$ & $-1$ 	& $4p^2q^3+pq^4$ \\ \hline 
	$-1$ & $-3$ 	& $4pq^4$ \\ \hline 
	$-3$ & $-3$	& $q^5$ \\ \hline 
	\hline 
\end{tabular}
\end{table}

\subsection{Blinking structures}

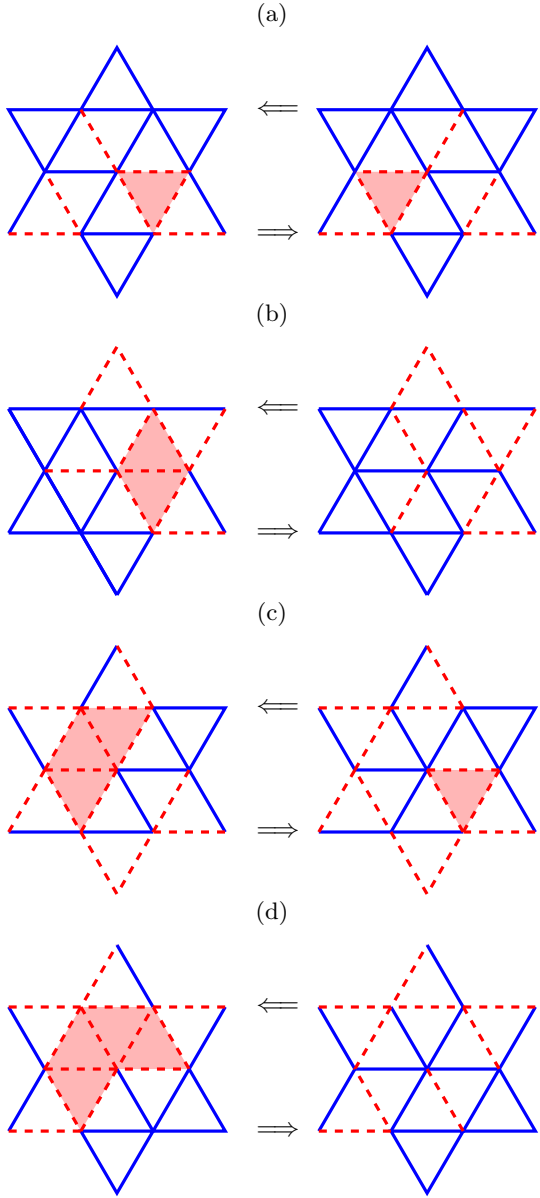
\begin{figure}[htbp]
\begin{subfigure}{.99\columnwidth}
\caption{\label{bli02}}
	\centering
\begin{tikzpicture}[scale=0.95]
\tkzDefPoint(1.5,0.866){A}
\tkzDefPoint(2.5,0.866){B}
\tkzDefPoint(2,0){C}
\tkzFillPolygon[red!30,opacity=0.95](A,B,C)
\draw[blue,very thick]  	(1,0) -- (2,0) -- (1.5,-.866) -- (1,0) -- (1.5,.866) -- (0.5,.866);
\draw[blue,very thick]  	(0.5,.866) -- (0,1.732) -- (3,1.732) -- (2.5,.866);
\draw[blue,very thick]  	(0,0) -- (1.5,2.598) -- (3,0);
\draw[blue,very thick]  	(1.5,.866) -- (2,1.732);
\draw[red,very thick,dashed]    (0,0) -- (1,0) -- (0.5,.866);
\draw[red,very thick,dashed]    (3,0) -- (2,0) -- (2.5,.866) -- (1.5,.866);
\draw[red,very thick,dashed]    (1,1.732) -- (2,0);
\node[right] at (3.3, 1.732) {$\Longleftarrow$};
\node[right] at (3.3, 0) {$\Longrightarrow$};
\end{tikzpicture}
\begin{tikzpicture}[scale=0.95]
\tkzDefPoint(1.5,0.866){A}
\tkzDefPoint(0.5,0.866){B}
\tkzDefPoint(1,0){C}
\tkzFillPolygon[red!30,opacity=0.95](A,B,C)
\draw[blue,very thick]          (1,0) -- (2,0) -- (1.5,-.866) -- (1,0);
\draw[blue,very thick]          (0.5,.866) -- (0,1.732) -- (3,1.732) -- (2.5,.866);
\draw[blue,very thick]          (0,0) -- (1.5,2.598) -- (3,0);
\draw[blue,very thick]    	(1,1.732) -- (2,0);
\draw[blue,very thick]    	(1.5,.866) -- (2.5,.866);
\draw[red,very thick,dashed]    (0,0) -- (1,0) -- (0.5,.866);
\draw[red,very thick,dashed]    (3,0) -- (2,0) -- (2.5,.866);
\draw[red,very thick,dashed]    (1,0) -- (1.5,.866) -- (0.5,.866);
\draw[red,very thick,dashed]    (1.5,.866) -- (2,1.732);
\end{tikzpicture}
\end{subfigure}
\begin{subfigure}{.99\columnwidth}
\caption{\label{bli22}}
	\centering
\begin{tikzpicture}[scale=0.95]
\tkzDefPoint(1.5,.866){A}
\tkzDefPoint(2,1.732){B}
\tkzDefPoint(2.5,.866){C}
\tkzDefPoint(2,0){D}
\tkzFillPolygon[red!30,opacity=0.95](A,B,C,D)
\draw[blue,very thick] (1,0) -- (1.5,.866) -- (1,1.732);
	\draw[blue,very thick] (0,1.732) -- (3,1.732);
        \draw[blue,very thick] (1.5,-.866) -- (0,1.732);
        \draw[blue,very thick] (0,0) -- (2,0);
        \draw[blue,very thick] (1.5,-.866) -- (0,1.732);
        \draw[blue,very thick] (0,0) -- (.5,.866);
        \draw[blue,very thick] (1.5,-.866) -- (2,0);
        \draw[blue,very thick] (2.5,.866) -- (3,0);
        \draw[blue,very thick] (.5,.866) -- (1,1.732);
\draw[red,very thick,dashed] (.5,.866) -- (2.5,.866);
\draw[red,very thick,dashed] (2,0) -- (1.5,.866) -- (2,1.732);
	\draw[red,very thick,dashed] (1,1.732) -- (1.5,2.598) -- (2.5,.866);
	\draw[red,very thick,dashed] (3,0) -- (2,0) -- (3,1.732);
\node[right] at (3.3, 1.732) {$\Longleftarrow$};
\node[right] at (3.3, 0) {$\Longrightarrow$};
\end{tikzpicture}
\begin{tikzpicture}[scale=0.95]
\draw[blue,very thick] (.5,.866) -- (2.5,.866);
\draw[blue,very thick] (2,0) -- (1.5,.866) -- (2,1.732);
	\draw[blue,very thick] (0,1.732) -- (3,1.732);
        \draw[blue,very thick] (0,0) -- (2,0);
        \draw[blue,very thick] (1.5,-.866) -- (0,1.732);
        \draw[blue,very thick] (0,0) -- (.5,.866);
        \draw[blue,very thick] (1.5,-.866) -- (2,0);
        \draw[blue,very thick] (2.5,.866) -- (3,0);
        \draw[blue,very thick] (.5,.866) -- (1,1.732);
\draw[red,very thick,dashed] (1,0) -- (1.5,.866) -- (1,1.732);
	\draw[red,very thick,dashed] (1,1.732) -- (1.5,2.598) -- (2.5,.866);
	\draw[red,very thick,dashed] (3,0) -- (2,0) -- (3,1.732);
\end{tikzpicture}
\end{subfigure}
\begin{subfigure}{.99\columnwidth}
\caption{\label{bli12}}
\centering
\begin{tikzpicture}[scale=0.95]
\tkzDefPoint(1,0){A}
\tkzDefPoint(.5,0.866){B}
\tkzDefPoint(1,1.732){C}
\tkzDefPoint(2,1.732){D}
\tkzFillPolygon[red!30,opacity=0.95](A,B,C,D)
	\draw[blue,very thick] (0,0) -- (2,0);
	\draw[blue,very thick] (3,0) -- (2.5,.866) -- (3,1.732) -- (2,1.732) -- (2.5,.866);
	\draw[blue,very thick] (1,1.732) -- (1.5,2.598);
	\draw[blue,very thick] (.5,.866) -- (0,1.732);
\draw[blue,very thick] (1.5,.866) -- (2.5,.866);
\draw[blue,very thick] (1.5,.866) -- (2,0);
\draw[red,very thick,dashed] (1,0) -- (2,1.732);
\draw[red,very thick,dashed] (1.5,.866) -- (.5,.866);
\draw[red,very thick,dashed] (1.5,.866) -- (1,1.732);
	\draw[red,very thick,dashed] (1,0) -- (1.5,-.866) -- (2,0) -- (3,0);
	\draw[red,very thick,dashed] (2,0) -- (2.5,.866);
	\draw[red,very thick,dashed] (1,1.732) -- (2,1.732);
	\draw[red,very thick,dashed] (1,1.732) -- (.5,0.866);
	\draw[red,very thick,dashed] (2,1.732) -- (1.5,2.598);
	\draw[red,very thick,dashed] (0,0) -- (.5,.866) -- (1,0);
	\draw[red,very thick,dashed] (0,1.732) -- (1,1.732);
\node[right] at (3.3, 1.732) {$\Longleftarrow$};
\node[right] at (3.3, 0) {$\Longrightarrow$};
\end{tikzpicture}
\begin{tikzpicture}[scale=0.95]
\tkzDefPoint(1.5,.866){A}
\tkzDefPoint(2.5,.866){B}
\tkzDefPoint(2,0){C}
\tkzFillPolygon[red!30,opacity=0.95](A,B,C)
	\draw[blue,very thick] (0,0) -- (2,0);
	\draw[blue,very thick] (3,0) -- (2.5,.866) -- (3,1.732) -- (2,1.732) -- (2.5,.866);
	\draw[blue,very thick] (1,1.732) -- (1.5,2.598);
	\draw[blue,very thick] (.5,.866) -- (0,1.732);
\draw[blue,very thick] (1,0) -- (2,1.732);
\draw[blue,very thick] (1.5,.866) -- (.5,.866);
\draw[blue,very thick] (1.5,.866) -- (1,1.732);
\draw[red,very thick,dashed] (1.5,.866) -- (2.5,.866);
\draw[red,very thick,dashed] (1.5,.866) -- (2,0);
        \draw[red,very thick,dashed] (1,0) -- (1.5,-.866) -- (2,0) -- (3,0);
        \draw[red,very thick,dashed] (2,0) -- (2.5,.866);
	\draw[red,very thick,dashed] (1,1.732) -- (2,1.732);
	\draw[red,very thick,dashed] (2,1.732) -- (1.5,2.598);
	\draw[red,very thick,dashed] (0,0) -- (.5,.866) -- (1,0);
	\draw[red,very thick,dashed] (0,1.732) -- (1,1.732);
	\draw[red,very thick,dashed] (0.5,.866) -- (1,1.732);
\end{tikzpicture}
\end{subfigure}
\begin{subfigure}{.99\columnwidth}
\caption{\label{bli32}}
	\centering
\begin{tikzpicture}[scale=0.95]
\tkzDefPoint(1,0){A}
\tkzDefPoint(.5,.866){B}
\tkzDefPoint(1,1.732){C}
\tkzDefPoint(2,1.732){D}
\tkzDefPoint(2.5,.866){E}
\tkzDefPoint(1.5,.866){F}
\tkzFillPolygon[red!30,opacity=0.95](A,B,C,D,E,F)
\draw[blue,very thick] (1.5,.866) -- (2,0);
	\draw[blue,very thick] (2,0) -- (1,0) -- (1.5,-.866) -- (2,0) -- (2.5,.866) -- (3,0) -- (2,0);
	\draw[blue,very thick] (0,0) -- (.5,.866) -- (0,1.732);
	\draw[blue,very thick] (2.5,.866) -- (3,1.732);
	\draw[blue,very thick] (2,1.732) -- (1.5,2.598);
\draw[red,very thick,dashed] (1.5,.866) -- (1,1.732);
\draw[red,very thick,dashed] (.5,.866) -- (2.5,.866);
\draw[red,very thick,dashed] (1,0) -- (2,1.732);
	\draw[red,very thick,dashed] (0,0) -- (1,0) -- (.5,.866) -- (1,1.732) -- (2,1.732) -- (2.5,.866);
	\draw[red,very thick,dashed] (0,1.732) -- (1,1.732) -- (1.5,2.598);
	\draw[red,very thick,dashed] (2,1.732) -- (3,1.732);
\node[right] at (3.3, 1.732) {$\Longleftarrow$};
\node[right] at (3.3, 0) {$\Longrightarrow$};
\end{tikzpicture}
\begin{tikzpicture}[scale=0.95]
\draw[blue,very thick] (1.5,.866) -- (1,1.732);
\draw[blue,very thick] (.5,.866) -- (2.5,.866);
\draw[blue,very thick] (1,0) -- (2,1.732);
	\draw[blue,very thick] (2,0) -- (1,0) -- (1.5,-.866) -- (2,0) -- (2.5,.866) -- (3,0) -- (2,0);
	\draw[blue,very thick] (0,0) -- (.5,.866) -- (0,1.732);
	\draw[blue,very thick] (2.5,.866) -- (3,1.732);
	\draw[blue,very thick] (2,1.732) -- (1.5,2.598);
\draw[red,very thick,dashed] (1.5,.866) -- (2,0);
	\draw[red,very thick,dashed] (0,0) -- (1,0) -- (.5,.866) -- (1,1.732) -- (2,1.732) -- (2.5,.866);
	\draw[red,very thick,dashed] (0,1.732) -- (1,1.732) -- (1.5,2.598);
	\draw[red,very thick,dashed] (2,1.732) -- (3,1.732);
\end{tikzpicture}
\end{subfigure}
\caption{\label{bli}Examples of configurations which follow each other in a cycle of length two, for the dilution parameter $f=0$.}
\end{figure}

When observing the time evolution of the lattice ruled by Eq.~\eqref{rule} after more than ten time steps, we notice some structures which are blinking with period two. These structures are most ubiquitous for $f$ positive but small. Yet, some of them appear already for $f=0$. In Fig. \ref{bli} some of them are shown, in two stages which appear alternately. As far as we can state, the automaton belongs to the second class in Wolfram classification \cite{Wolfram-2002}.

\section{\label{sec-mod}Model evaluation of $U(f)$}

\begin{figure}[htbp]
\centering
\begin{tikzpicture}[scale=1.5]
\draw[black,very thick] (0,0) -- (1,0);
\node[below] at (0.5, 0) {$S_b$};
\draw[black,very thick] (0,0) -- (0.5,0.866);
\node[left] at (0.25, 0.433) {$S_a$};
\draw[black,very thick] (1,0) -- (0.5,0.866);
\draw[black,very thick] (-0.5,0.866) -- (0.5,0.866);
\draw[black,very thick] (-0.5,0.866) -- (0,0);
\draw[black,very thick] (1.5,0.866) -- (0.5,0.866);
\draw[black,very thick] (1.5,0.866) -- (1,0);
\draw[black,very thick] (0.5,-0.866) -- (0,0);
\draw[black,very thick] (0.5,-0.866) -- (1,0);
\node[right] at (0.75, 0.433) {$S_c$};
\node[] at (-0.5, 0.433) {$S_1$};
\node[] at (0, 1.1) {$S_2$};
\node[] at (1, 1.1) {$S_3$};
\node[] at (1.5, 0.433) {$S_4$};
\node[] at (1., -0.433) {$S_5$};
\node[] at (0., -0.433) {$S_6$};
\end{tikzpicture}
\caption{\label{ren2}A central triad (links $S_a$, $S_b$, $S_c$) in a frozen neighbourhood (links $S_1$--$S_6$).}
\end{figure}
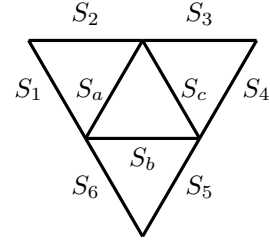

To address the dependence of energy $U$ on $f$, we consider a triad with links $S_a$, $S_b$, $S_c$ in a frozen neighbourhood of three neighbouring triads, as shown in Fig.~\ref{ren2}. The environment is formed by links $S_1$--$S_6$ between three central nodes and three other nodes which are nearest neighbours of two of the central ones. As shown in Fig.~\ref{roro}, the initial probability 0.5 of positive links remain unchanged by the time evolution. Then, each configuration of the six links can appear with the same probability. We can ask, which configurations out of $2^6=64$ states of the environment stabilise an unbalanced configuration? This question is considered separately for both unbalanced states. For three negative links in the central triad $S_a=S_b=S_c=-1$, the configurations which stabilise it are listed in Table~\ref{t1}. For an exemplary unbalanced configuration $S_a=-1$, $S_b=S_c=+1$ the configurations which stabilise it are listed in Table~\ref{t2}. In both cases, the number of such configurations is eight out of 64. On the contrary, all balanced configurations remain unchanged during the time evolution, for any of 64 states of their environment. This is an indication, that an initially unbalanced triad remains unbalanced with probability $\frac{1}{8}$, and switches to be balanced with probability $\frac{7}{8}$. On the other hand, initially balanced states of a triad remain balanced. Mean value of energy coming from initially balanced states is $-1$, and from initially unbalanced states is $\frac{7}{8}\cdot(-1)+\frac{1}{8}\cdot(+1)=-\frac{3}{4}$, which gives $\langle U\rangle=-\frac{7}{8}$ in the average. We add that for $f=0$, blinking configurations are very rare. Yet they exist: we show the observed ones in Fig.~\ref{bli}.

\begin{figure}[hptb]
\begin{subfigure}{.99\columnwidth}
\caption{\label{onenblia}}
\centering
\begin{tikzpicture}[scale=1.1]
\tkzDefPoint(3,0){A}
\tkzDefPoint(4,0){B}
\tkzDefPoint(3.5,.866){C}
\tkzFillPolygon[red!30,opacity=0.95](A,B,C)
\draw[blue,very thick] (0,0)--(1,0)--(.5,.866)--(-.5,.866);
\draw[red,very thick,dashed] (-.5,.866)--(0,0)--(.5,.866);
\node[right] at (1.3, .666) {$\Longleftarrow$};
\node[right] at (1.3, .2) {$\Longrightarrow$};
\draw[blue,very thick] (2.5,.866)--(3.5,.866);
\draw[red,very thick,dashed] (2.5,.866)--(3,0)--(3.5,.866)--(4,0)--(3,0);
\end{tikzpicture}
\end{subfigure}
\begin{subfigure}{.99\columnwidth}
\caption{\label{onenblib}}
\centering
\begin{tikzpicture}[scale=1.1]
\draw[blue,very thick] (1,0)--(.5,.866)--(-.5,.866)--(0,0)--(.5,.866);
\draw[red,very thick,dashed] (0,0)--(1,0);
\node[right] at (1.3, .666) {$\Longleftarrow$};
\node[right] at (1.3, .2) {$\Longrightarrow$};
\draw[blue,very thick] (4,0)--(3,0)--(3.5,.866)--(2.5,.866)--(3,0); 
\draw[red,very thick,dashed] (4,0)--(3.5,.866);
\end{tikzpicture}
\end{subfigure}
\caption{Examples of configurations which follow each other in a cycle of length two, between states ($-3$) and (+1) (\ref{onenblia}) and two states (+1) (\ref{onenblib}). The blinking triad has only one neighbour.}
\label{onenbli}
\end{figure}
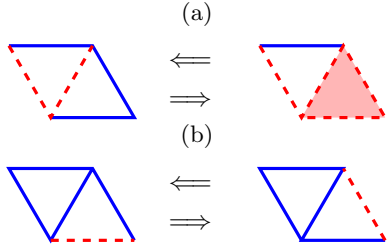

For $f>0$, we should take into account finite probabilities that a triad has less than three neighbouring complete triads. If the number of neighbours is zero, each triad becomes balanced and the mean energy is $-1$. We checked by inspection that also each triad with two neighbours always ends up in a balanced state. However triads with one neighbour, if initially unbalanced, remain unbalanced for half of configurations of their neighbours. It is interesting to note that such a triad is blinking forever between the states ($-3$) and (+1) or between two different states (+1), (+1). This evolution is shown in Fig.~\ref{onenbli}. Summarising, the mean energy of triads with one neighbour is $\langle U\rangle=\frac{1}{2}\cdot(-1)+\frac{1}{2}\cdot\left[\frac{1}{2}\cdot(+1)+\frac{1}{2}\cdot(-1)\right]=-\frac{1}{2}$.

\begin{figure}[htbp]
\centering
\includegraphics[width=.80\columnwidth]{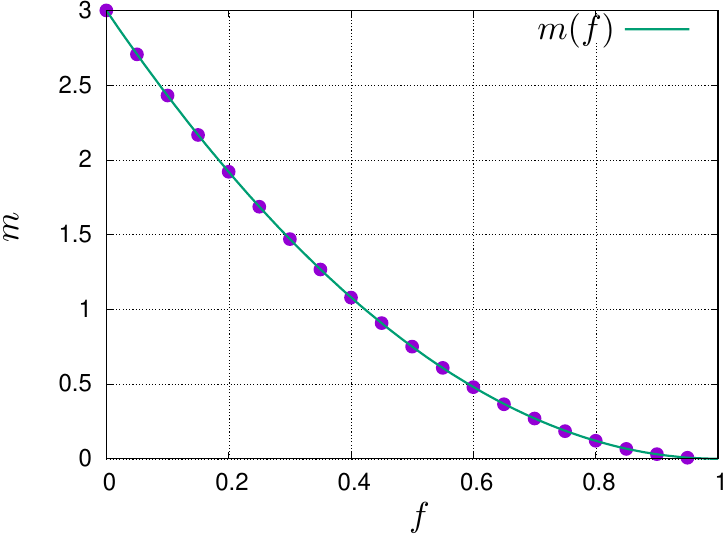}
\caption{\label{numtria}The mean number $m$ of triads which are nearest neighbours of a given triad, as dependent on the dilution parameter $f$.}
\end{figure}

Above we have calculated the energies $\langle U(k)\rangle$ as dependent on the number $k$ of neighbours: $\langle U(0)\rangle=\langle U(2)\rangle=-1$, $U(1)=-\frac{1}{2}$, and $\langle U(3)\rangle=-\frac{7}{8}$. It remains to find the probability distribution $R(k)$, as dependent on the dilution parameter $f$. These are taken from the Bernoulli distribution. A triad has a neighbour with probability $h=(1-f)^2$, hence $R(0)=(1-h)^3$, $R(1)=3h(1-h)^2$, $R(2)=3h^2(1-h)$, and $R(3)=h^3$. The resulting plot of the `average over averages' 
\begin{equation}
\bar U = \sum_{k=0}^3 R(k)\langle U(k)\rangle
\label{sumeq}
\end{equation}
is shown on Fig.~\ref{ufin} as a solid line, together with the result of the simulations.
The same distribution gives the mean value of the number of complete triads which are nearest neighbours of a given triad. The results of simulations
and of the analytical formula are shown in Fig.~\ref{numtria} as dots and continuous line, respectively.

In this calculation, we ignore the
{rule}
that non-balanced triads are never neighbours of each other. Also, it is possible to introduce the observed weights of triad types into the calculations, and it is likely that the accuracy is improved. However, the purpose of the analytical approach reported in this section is not to imitate the results of simulations, but rather to check which results of the simulations can be reproduced with simple assumptions.

\begin{table}[htbp]
\caption{\label{t1}The states of the frozen neighbourhood (links $S_1$--$S_6$ in Fig.~\ref{ren2}) which stabilise the unbalanced state $S_a=S_b=S_c=-1$.}
\centering
\begin{tabular}{|c|c|c|c|c|c|c|c|c|}
\hline
\hline 
$S_1$ & $S_2$ & $S_3$ & $S_4$ & $S_5$ & $S_6$ \\
\hline
\hline
$-1$&+1&$-1$&+1&$-1$&+1\\ 
\hline 
$-1$&+1&$-1$&+1&+1&$-1$\\ 
\hline 
$-1$&+1&+1&$-1$&$-1$&+1\\ 
\hline 
$-1$&+1&+1&$-1$&+1&$-1$\\ 
\hline 
+1&$-1$&$-1$&+1&$-1$&+1\\ 
\hline 
+1&$-1$&$-1$&+1&+1&$-1$\\ 
\hline 
+1&$-1$&+1&$-1$&$-1$&+1\\ 
\hline 
+1&$-1$&+1&$-1$&+1&$-1$\\ 
\hline 
\hline 
\end{tabular}
\end{table}

\begin{table}[htbp]
\caption{\label{t2}The states of the frozen neighbourhood (links $S_1$--$S_6$ in Fig.~\ref{ren2}) which stabilise the unbalanced state $S_a=-1$, $S_b=S_c=+1$.}
\centering
\begin{tabular}{|c|c|c|c|c|c|c|c|c|}
\hline
\hline 
$S_1$ & $S_2$ & $S_3$ & $S_4$ & $S_5$ & $S_6$ \\
\hline
\hline
$-1$&+1&+1&+1&+1&+1\\ 
\hline 
+1&$-1$&+1&+1&+1&+1\\ 
\hline 
$-1$&+1&$-1$&$-1$&$-1$&$-1$\\ 
\hline 
+1&$-1$&$-1$&$-1$&$-1$&$-1$\\ 
\hline 
$-1$&+1&$-1$&$-1$&+1&+1\\ 
\hline 
+1&$-1$&$-1$&$-1$&+1&+1\\ 
\hline 
$-1$&+1&+1&+1&$-1$&$-1$\\ 
\hline 
+1&$-1$&+1&+1&$-1$&$-1$\\ 
\hline 
\hline 
\end{tabular}
\end{table}

\section{Discussion}

The calculated correlations between neighbouring balanced triads indicate, that both the standard correlations and the correlations with extracted part caused by common links of neighbouring triads are further enhanced by the time evolution, if only the dilution remains limited. This can be interpreted as a kind of attraction between balanced triads. We interpret this effect as an analogy to the gathering of positive links within internally friendly groups. That gathering is known to appear in structurally balanced states of complete graphs. However, on the contrary to the latter, here the balance is never complete.

As expected, the energy decreases in time. Its dependence on the dilution factor $f$ shows a maximum, shown in Fig.~\ref{ufin}. Its shape is qualitatively reproduced by the approximate formulae, obtained with assumption that it is only a triad  which evolves, and not its neighbours. We coined a term `frozen neighbourhood’. This approximation systematically underestimates the probability of non-balanced states. Indeed, an observation of local configurations indicates that more than often, the evolving part of lattice contains more than one triad. Including this effect to the evaluation of $\langle U(f)\rangle$ should improve the accuracy of the theoretical result. However, the approximation is good enough to interpret the observed maximum of $\langle U(f)\rangle$ as the result of the configurations of a triad with only one neighbour.

As noted above, if the concentration of positive\-/\-nega\-tive links is initially 0.5, this value holds throughout the time evolution.
This allows to consider the process as just {a} reordering the signs of links.
Recall that for complete graphs the final {balanced} state is usually equivalent to two mutually hostile but internally friendly groups.
The triads (+1) destroy this coherence, as the friendly relation is not transitive there.
However---again by inspection of a final state---a path through positive links spans over most nodes.
The same is true for a path through negative links.
{Actually, the bond percolation limit in the triangular lattice is $p_c=2\sin(\pi/18)$, i.e. about 0.347} \cite{bookDS}{.}
We note that although the number of unbalanced triads is reduced during the simulation, the ordering of links specific for the balance does not destroy the largest clusters
{of positive  (negative) links.}
{This is because the removal of positive or negative links leaves a half of links in both cases, and their concentration $0.5$ is still above the percolation limit.}

{We note that for $f>1-p_c$, most of triads are mutually isolated, and the approach proposed in the preceding section works quite well, as seen in Fig.}~\ref{ufin}{. For smaller value of $f$, interaction of neighboring triads influences the final energy; in Section}~\ref{sec-mod}{, this interaction is only approximated and the results are different.}

The automaton presented here is---up to our know\-ledge\----the first local formulation of the process leading to Heider balance. It allows to investigate the dependence of stability and dynamics on the network topology, if only the clustering coefficient is not too small. As noted for example in Ref.~\cite{PhysRevE.68.036122}, high values of the clustering coefficient are characteristic for social networks. It is obvious that the problem of structural balance finds applications right there. On the other hand, we have identified some dynamic (blinking) configurations, which is also a rare opportunity in models of the structural balance. For an exception we refer to Ref.~\cite{Gawronski}, where oscillations have been found in a system of an isolated triad with asymmetric links.

{The model time evolution considered here can be generalized by adding an external field, which promotes positive or negative links.
This transformation is more natural if we substitute links by spins equal to $\pm 1$. The interaction between the links is then converted
into a three-spin term. The links between the interacting spins form a new lattice, the so-called line graph} \cite{Wierman2014,MankaKrason2010118}{, where two spins are connected
if the links in the initial lattice meet in a common node. The site percolation threshold of this line graph is the same as the bond percolation threshold
of the triangular lattice} \cite{Wierman2014}{. In such a spin system,
an external field is coupled to the spins in the same way as in the Zeeman term in the Ising model} \cite{Huang}{.
Now, suppose that we perform a hysteresis experiment in thought, starting from a balanced configuration with boundaries, as in Fig.}~\ref{lattice}{. 
Each spin is connected with two pairs of neighbors, as in Eq.}~\eqref{U}{. The applied field which promotes friendly relation is expected to switch
the hostile relations when equal to +2 (i.e. the number of neighboring pairs). Now consider the paradise (saturated) state and the negative applied
field which promotes hostile relations. Again, in the balanced state the links are expected to be modified at the field equal to $-2$. This shows that the area of the hysteresis loop is non-zero. In other words, the dynamics is dissipative.}

\section*{{Acknowledgements}}
{We are grateful to anonymous Referees for their valuable comments.}

\bibliographystyle{apsrev4-2-titles}
\bibliography{../../../BIBLIOGRAPHY/heider,../../../BIBLIOGRAPHY/percolation,../../../BIBLIOGRAPHY/ca,../../../BIBLIOGRAPHY/km,this}
\end{document}